\pdfoutput=1 

\documentclass[journal]{IEEEtran}
 \usepackage{amsmath}
\usepackage{bm}
\usepackage{cite} 
\usepackage[pdftex]{graphicx}% Include figure files
\usepackage{epstopdf}
\usepackage{dcolumn}% Align table columns on decimal point
\usepackage{bm}% bold math
\usepackage{float}
\usepackage{color}

\ifCLASSINFOpdf
\else
\fi

\begin{document}
\title{Engineering Nanowire n-MOSFETs at $L_{g}<$8 nm}

\author{ Saumitra Mehrotra, SungGeun Kim,Tillmann Kubis, Michael Povolotskyi, Mark Lundstrom and Gerhard Klimeck\\%
\thanks{Manuscript received on xxx. Authors acknowledge the financial support from SRC, FCRP-MSD and NSF. Computational resourses provided by nanoHUB.org is also acknowledged. S. Mehrotra, S.G. Kim, M. Povolotskyi, T. Kubis, M. Lundstrom and G. Klimeck are with the Network for Computational Nanotechnology and the Department of Electrical and Computer Engineering, Purdue University, West Lafayette, IN-47906, USA.(e-mail:saumitra.r@gmail.com).}}%
\markboth{IEEE Electron Device Letter,~Vol.~1, No.~1,~xxx~2012}{Mehrotra \MakeLowercase{\textit{et al.}}:Engineering Nanowire n-MOSFETs at $L_{g}<$8 nm}


\markboth{}%
{}

\maketitle

\begin{abstract}
 As metal-oxide-semiconductor field-effect transistors (MOSFET) channel lengths ($L_{g}$) are scaled to lengths shorter than $L_{g}$$<$8 nm  source-drain tunneling starts to become a major performance limiting factor. In this scenario a heavier transport mass can be used to limit source-drain (S-D) tunneling. Taking InAs and Si as examples, it is shown that different  heavier transport masses can be engineered using strain and crystal orientation engineering. Full-band extended device atomistic quantum transport simulations are performed for nanowire MOSFETs at $ L_{g}$$<$8 nm  in both ballistic and incoherent scattering regimes. In conclusion, a heavier transport mass can indeed be advantageous in improving ON state currents in ultra scaled nanowire MOSFETs.
\end{abstract}

\begin{IEEEkeywords}
Source-Drain tunneling, nanowire, Si, InAs, strain, quantum transport, tight-binding (TB) approach.
\end{IEEEkeywords}

% \fi
\IEEEpeerreviewmaketitle

\section{Introduction}
\IEEEPARstart{S}{caling} of complementary metal-oxide-semiconductor (CMOS) technology for the past fourty years has led to current device technology with channel lengths well below 30 nm \cite{22nm}. The International Technology Roadmap for Semiconductors (ITRS) predicts MOSFET channel lengths to be less than 8 nm in ten years \cite{ITRS}.  In this extremely scaled regime, MOSFETs will suffer from excessive source - drain (S-D) tunneling, making it hard to turn off the device \cite{Mathieu2011,Wang2002}. Nanowire based MOSFETs have emerged as promising candidates for future scaling as they offer the best electrostatic gate control over the channel \cite{Appenzeller2008}. Nanowire MOSFETs made from high mobility III-V materials are being projected as the future of microelectronics \cite{Alamo2011}. However, it remains an open question how practical it would be to scale MOSFETs to channel lengths below 8 nm. In such extremely scaled regime of operation, it becomes clear that the most critical aspect is to maintain good sub-threshold characteristics. 

The typical transistor approach prefers a light transport mass corresponding to high carrier velocities and a heavy confinement mass for higher quantum capacitance ($C_{q}$) \cite{Fischetti2007}. A light transport mass, however, leads to an increased source drain tunneling and can lead to degraded OFF state characteristics \cite{Sylvia2012}. A heavy transport mass can limit S-D tunneling but it also means lower channel mobility or degraded ON state characteristics. This situation naturally leads to a trade-off and begs the question - what transport mass will work the best for ultra scaled channel MOSFETs and can it be engineered?

 This work shows that different conduction band (CB) minima masses can be obtained in nanowire channels using strain and orientation engineering and how these changes will affect the transistor performance. In a Si nanowire, due to quantum confinement, the six bulk $\Delta$ valleys  rearrange themselves in energy, with each equivalent set of $\Delta$ valleys forming its own separate "energy ladder" . Furthermore, the different sets of "energy ladders" can be "rearranged' in energy space using uniaxial strain leading to different band edge transport masses (Fig. 1). InAs as a channel material is also considered in this study as a candidate for "high-mobility material". Full band quantum transport calculations are used to assess the ON state ballistic and electron-phonon scattering limited device performance for the different CB minima mass nanowire devices. The simulation results suggest that for scaling of the channel lengths below 8 nm, increasing the transport mass leads to better device performance.

\begin{figure}[t]
\centering
\includegraphics[scale=1.0]{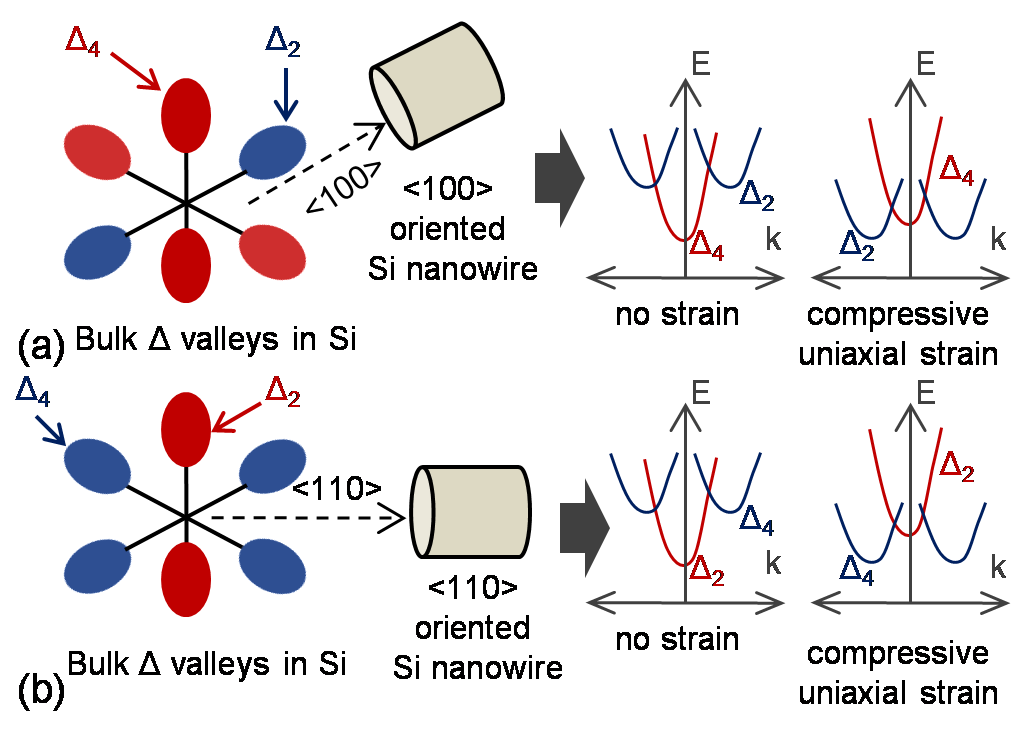}
 \caption{Schematic description of the formation of "energy ladders" or sub-bands in (a) $<$100$>$ and (b) $<$110$>$ oriented Si nanowire under no strain and compressive strain conditions. }\label{A}
\end{figure}

\begin{figure*}[t]
\centering
\includegraphics[scale=1.0]{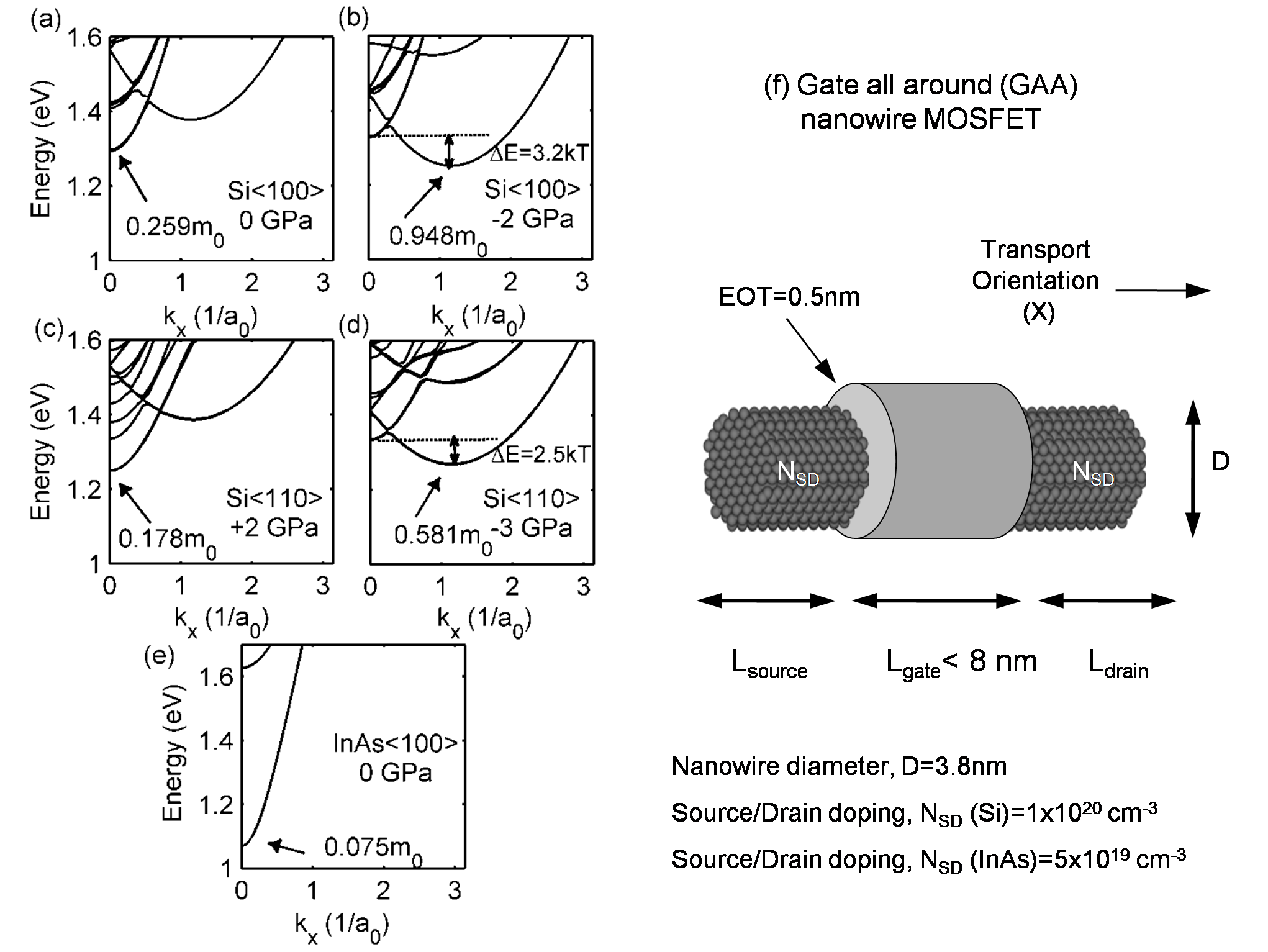}
 \caption{Calculated bandstructure for (a) $<$100$>$ oriented Si, (b) compressively strained $<$100$>$ oriented Si,  (c) tensile strained $<$110$>$ oriented Si, (d) compressively strained $<$110$>$ oriented Si and (e)  $<$100$>$ oriented InAs nanowire with 3.8 nm diameter. (f) Schematic view of the simulated device structure. }\label{A}
\end{figure*}

\section{Simulation Approach}
\subsection{Nanowire cases considered in this study}
 Five different nanowire MOSFET cases are studied in this paper. At sub-8 nm channel lengths, a smaller diameter is critical for an increased gate control.  At the same time, however, a smaller diameter leads to increased threshold voltage fluctuation due to process variations. Considering these issues, an optimal diameter, D=3.8 nm, is used for all the simulations \cite{Yu2010}. As a first step the different CB minina mass conditions that can be achieved  in Si and InAs nanowires are discussed.The applied stress values were chosen to be high enough such that the bottom-most band is  energetically at least 2kT lower than the next higher sub-band, while at the same time ensuring the stress values are experimentally achievable \cite{Thompson2006}.  
\subsubsection{Unstrained Si $<$100$>$}
 Fig. 2(a) shows the case for an unstrained \textless100\textgreater oriented Si nanowire where the $\Delta_{4}$ set of valleys form the CB minima . The CB minima effective mass is calculated to be 0.259$m_{o}$. 
\subsubsection{Compressive stressed Si $<$100$>$}
Unstrained Si $<$100$>$ nanowire exhibits $\Delta_{2}$ valleys that have a heavy transport mass but  lie higher in energy due to its lighter confinement mass. On application of compressive stress the heavy $\Delta_{2}$ valleys are pulled down in energy leading to CB minima mass of 0.948$m_{o}$ as shown in Fig. 2(b).
\subsubsection{Tensile stressed Si $<$110$>$}
 Fig. 2(c) shows the E-k relation for a tensile stressed $<$110$>$  oriented Si nanowire. This case is close to current n-MOS technology where tensile stress is used to reduce CB minima mass and increase the energy difference between the lighter $\Delta_{2}$ and heavier $\Delta_{4}$ valleys \cite{Thompson2006,22nm}. The CB minima mass for a $<$110$>$ oriented nanowire with an applied tensile stress of 2 GPa is calculated to be 0.178$m_{0}$.
\subsubsection{Compressive stressed Si $<$110$>$}
On applying a compressive stress along Si $<$110$>$ transport direction, heavier  $\Delta_{4}$ valleys are pulled down increasing the CB minima mass. As the stress type is reveresed to compressive 3 GPa, the CB minima mass increases to 0.581$m_{o}$  from 0.178$m_{o}$ as shown in Fig. 2(d).
\subsubsection{Unstrained InAs $<$100$>$}
 High mobility materials with high In\% are being actively researched as a post-Si channel material \cite{Alamo2011}. In light of this fact a $<$100$>$ oriented InAs channel is also considered in this study. Fig. 2(e) shows the E-k, with a calculated band edge mass of 0.075$m_{0}$.

\begin{figure*}[t]
\centering
\includegraphics[scale=1]{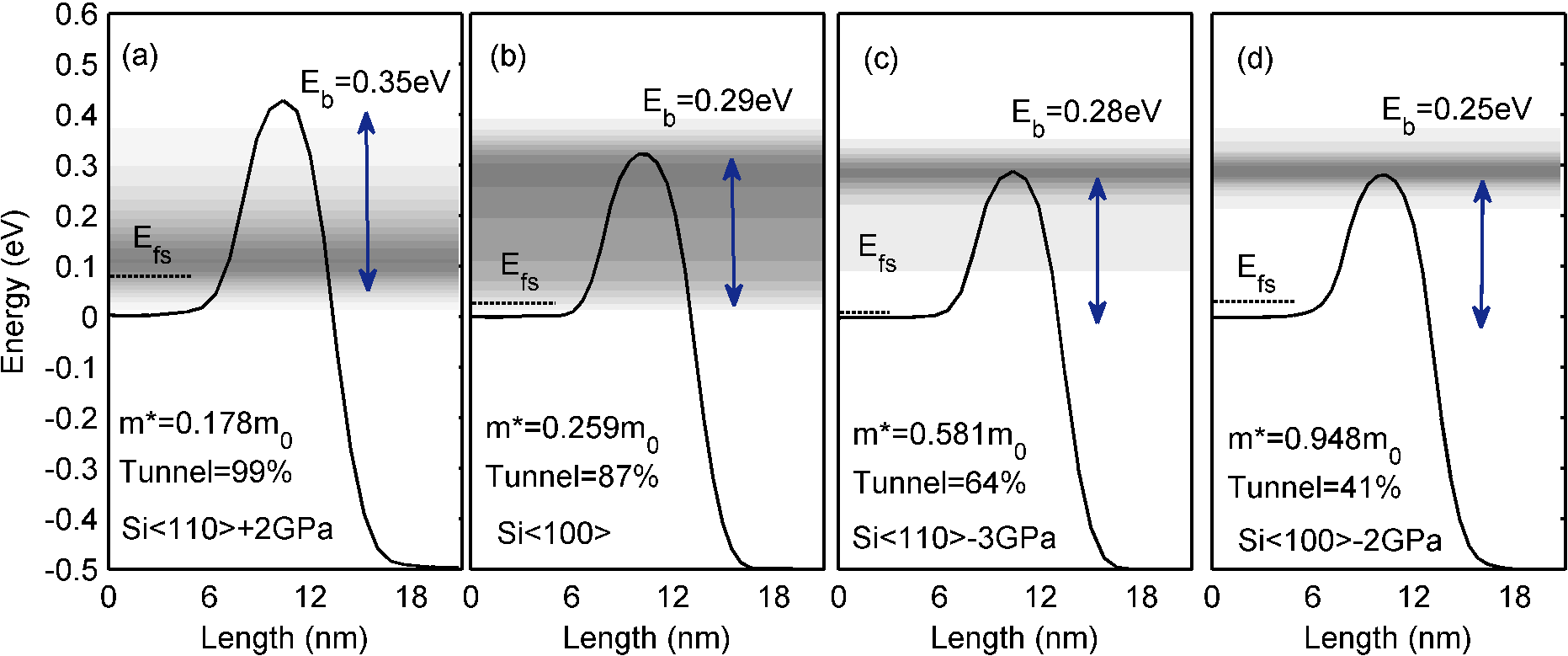}
 \caption{Illustration of the leakage current at OFF state ($I_{ds}$=0.1$\mu A/\mu m$) at $L_{g}$=5 nm. Normalized energy current spectrum is plotted for (a) Si $<$110$>$ +2GPa, (b) Si $<$100$>$, (c) Si $<$110$>$ -3GPa and (d) Si $<$100$>$ -2GPa nanowires cases. Due to S-D tunneling the barrier height needs to be much higher to achieve the desired $I_{OFF}$.} 
\end{figure*}

\subsection{Transport Simulation}

 Transport simulations are performed using a full-band quantum transport simulator based on the  $sp^{3}d^{5}s^{*}$ tight-binding model in the nearest neighbour approximation (without spin-orbit coupling). The atomistic Schrodinger-Poisson equations are solved self-consistently in the nonequilibrium Green function formalism at room temperature~\cite{OMEN,Si2004,InAs2002,newstrain}. The electron-phonon scattering is computed in the self-consistent Born approximation that couples the full electron and confined phonon spectra \cite{OMENphonon}. An effective oxide thickness (EOT) equal to 0.5 nm is used for all the simulations \cite{ITRS}. The oxide
layers are treated as perfect insulator and hard wall boundary conditions are applied to the surface Si atoms. A source/drain doping level of 1$\times$$10^{20}/cm^{3}$ is assumed for Si nanowires while 5$\times$$10^{19}/cm^{3}$ is the doping level for the InAs nanowire.

% The five different nanowire cases considered in this paper are shown in Fig. 1. Fig. 1 (a) shows the E-k relation for a 2 GPa tensile stressed \textless110\textgreater oriented Si nanowire. This case is close to current nMOS technology where tensile stress is used to reduce CB minima mass and increase the energy difference between the lighter $\Delta_{2}$ and heavier $\Delta_{4}$ valleys \cite{Thompson2006,22nm}. On applying a compressive stress along Si \textless110\textgreater transport direction, heavier  $\Delta_{4}$ valleys are pulled down increasing the CB minima mass  from 0.178$m_{o}$ to 0.581$m_{o}$ as shown in Fig. 1(b). 

Nanowire gate all around (GAA) n-MOSFETs are simulated for the five different CB minima mass cases (Figs. 2(a)-(e)) at channel lengths of $L_{g}=$3, 5 and 7 nm as shown in Fig. 2(f). The OFF state current ($I_{OFF}$) is set to $0.1\mu A/\mu m$ ($I_{OFF}$=$I_{DS}$ at $V_{GS}=0V$ and $V_{DS}=0.5V$), where the current has been normalized by the diameter \cite{ITRS}. Transfer characteristics of the different nanowire MOSFETs are compared at a supply voltage $V_{DS}$=0.5V. 

\section{S-D tunneling and subthreshold characteristics}

  The S-D tunneling is a quantum mechanical phenomenon that is fundamental and cannot be avoided. Fig. 3 illustrates the OFF state current profile for the different CB minima cases at $L_{g}$=5 nm simulated in the ballistic regime. For the heaviest mass case it can be seen that most of the current  flows over the barrier as opposed to the lightest mass case where most of the current quantum-mechanically tunnels through the barrier. The amount of current flowing below the potential barrier height ($E_{b}$) is the tunneling current ($I_{DS}^{tunnel}$). The ratio of $I_{DS}^{tunnel}$ to the total current  $I_{DS}^{total}$ or the $Tunnel\%$, increases as the CB minima mass becomes lighter. As a consequence of increased S-D tunneling, to achieve the desired OFF state current the barrier height needs to be raised much more for a light transport mass as compared to a heavy transport mass case. This leads to an increased SS in presence of S-D tunneling. 

\begin{figure}[h]
\centering
\includegraphics[scale=1]{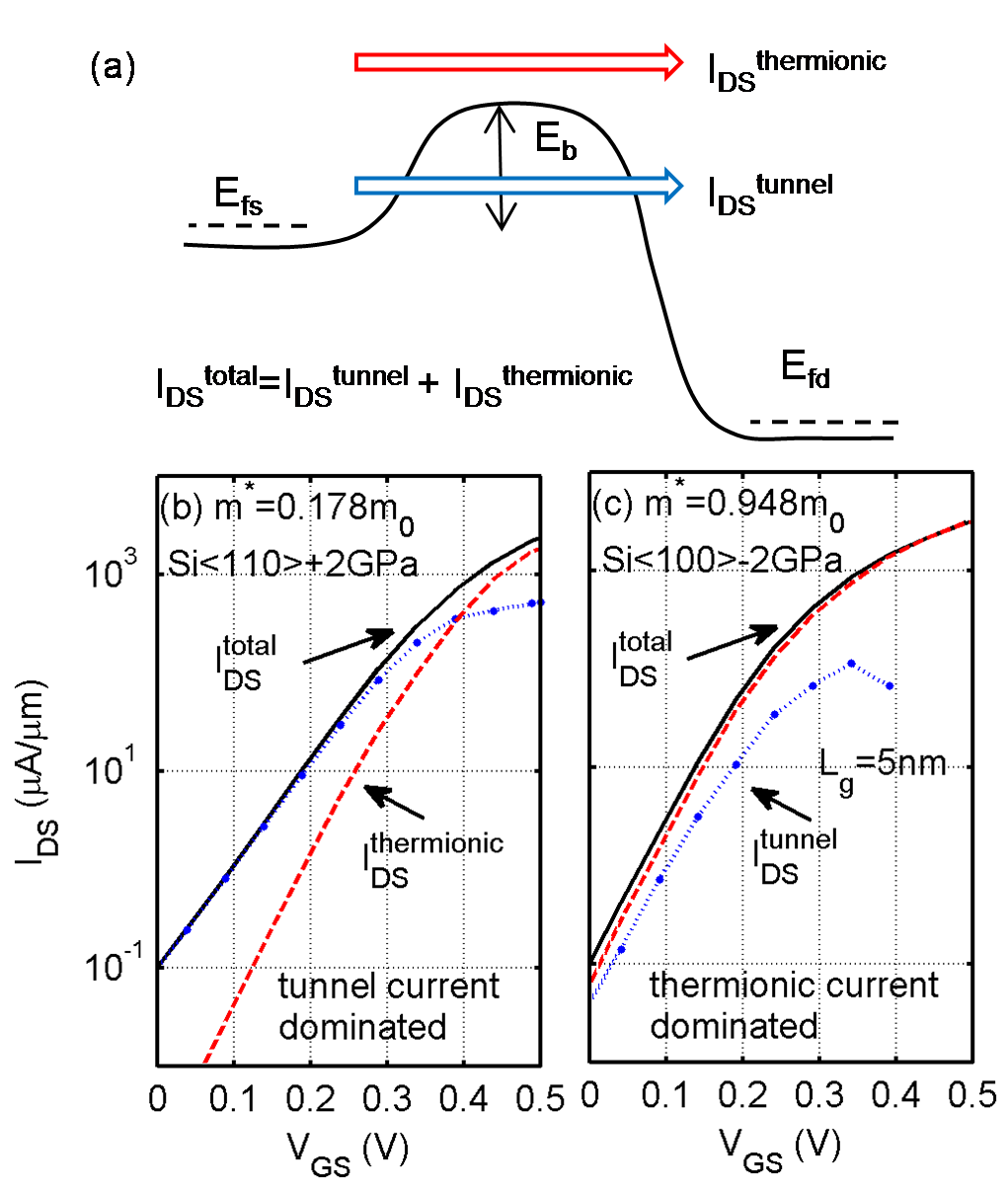}
\caption{ Calculated thermionic $I_{DS}^{thermionic}$ and tunnel component $I_{DS}^{tunnel}$ of the total current $I_{DS}^{total}$, for (a) Si $<$110$>$+2GPa and (b) Si $<$100$>$-2GPa nanowire cases. } 
\end{figure}

	Fig. 4(a) shows the tunnel component ($I_{DS}^{tunnel}$) along with the current flowing over the barrier or the thermionic component ($I_{DS}^{thermionic}$) during a MOSFET operation. Bias dependent current components for a light transport mass case (equivalent to Fig. 2(c)) is shown in Fig. 4(b) and a for heavy transport mass case (equivalent to Fig. 2(b)) is shown in Fig. 4(c), both at $L_{g}=5$ nm. The adverse effect of S-D tunneling in the presence of a light CB minima mass becomes clear as the subthreshold current is totally dominated by $I_{DS}^{tunnel}$. On the other hand, subthreshold slope is controlled by $I_{DS}^{thermionic}$ current for the heavy CB minima mass case, bringing it closer to the classical MOSFET operation where the electrostic gate control determines the subthreshold characteristics.
 
\begin{figure}[t]
\centering
\includegraphics[scale=1]{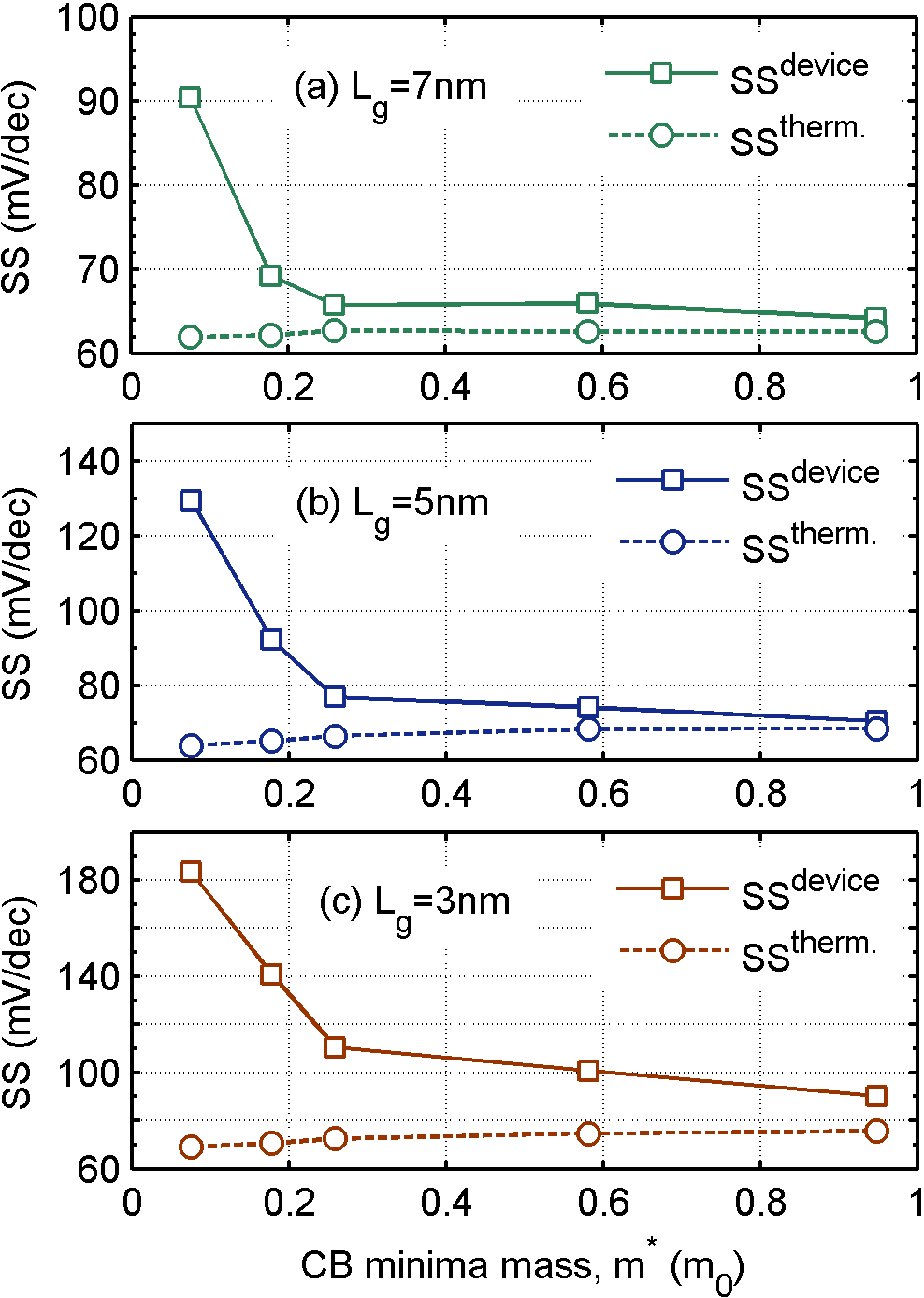}
\caption{ Subthreshold slope calculated from the total current ($SS^{device}$) and the thermionic current component ($SS^{therm.}$) for the different CB minima nanowire cases at (a) 7 nm , (b) 5 nm and (c) 3 nm channel lengths.}
\end{figure}

     The calculated subthreshold characteristics for the different CB minima cases at different channel lengths are shown in Fig. 5. The subthreshold slope calculated from $I_{DS}^{thermionic}$ is labelled as $SS^{therm.}$ while $SS^{device}$ is the actual subthreshold slope of the nanowire MOSFET calculated in presence of S-D tunneling. It can be readily observed that for light CB minima mass cases the $SS^{therm.}$ and $SS^{device}$ show a huge difference. As the CB minima mass increases, the $SS^{therm.}$ and $SS^{device}$ values begin to merge. This can be understood from the fact that with a heavy CB minima mass the electrostatic gate control determines the subthreshold slope. As the CB minima mass becomes lighter the S-D tunneling component begins to dominate. This fact also  underlines another important understanding that the lower limit of SS=60 mV/dec at room temperature does not hold anymore in presence of S-D tunneling. With the best of gate control, the amount of S-D tunneling will set the lower limit on the achievable SS and that will be more than 60mV/dec.

%The percent current that is tunneling through the barrier is a relevant metric to understand the extent of S-D tunneling. In the limiting condition of SS being solely degraded due to S-D tunneling as shown in Fig. 3(a), a simple analytical relation can be determined to relate the percent of the tunneling current ($Tunnel\%$) at the OFF state to the SS of that device  (Eq. (1)).
% \begin{eqnarray}
%$Tunnel\%=100\times \left( 1-{ \frac{I_{vt}}{I_{OFF}}}^{1-\frac{SS^{tunnel}}{60}} \right)$
%Tunnel\%=100\times  \left[ 1- \left(\frac{I_{OFF}}{I_{vth}}\right)^{\left(\frac{SS^{tunnel}}{60}-1\right)} \right]
%\end{eqnarray}

%Here, ${SS^{tunnel}}$ is the SS of the device in presence of tunneling current. $I_{Vth}$ is the current close to threshold voltage, which is labelled as $V_{th}$. Fig. 3(b) plots the analytical relation presented in Eq. (1) assuming $I_{OFF}$=$0.1 \mu A/\mu m$ and $I_{vth}$=$100 \mu A/\mu m$. Interestingly, the relation between $Tunnel\%$ and the subthreshold slope is directly but non-linearly related. Still the comparison of the analytical expression with the numerically calculated values provide a good qualitative agreement. Since the nanowire case with $L_{g}=7 nm$ is expected to have the best gate control, it provides the closest agreement with the analytical expression.
%The extracted $Tunnel\%$ at the OFF state and the corresponding SS are shown in Figs. 4(a) and 4(b) for the different channel length and CB minima effective mass cases. As expected a heavier CB minima mass nanowire device exhibits a smaller tunnneling current as well as a lower SS.

 It should be noted that, for different channel lengths the $SS^{therm.}$ is calculated to be $\sim$ 62 mV/dec at $L_{g}=7$ nm to  $\sim$ 75 mV/dec at $L_{g}=3$ nm. This fact highlights the strong electrostatic gate-control that can be achieved using a nanowire geometry. A light transport mass, however, cancels out any advantage because of increaed S-D tunneling. This is a crucial understanding that even with the best of gate electrostatics  the so called "high-mobility" or low mass materials may fail to perform as channel lengths are scaled to sub-8 nm dimensions.

\section{ON-state performance}

The calculated ballistic $I_{ON}$ currents ($I_{ON}$=$I_{DS}$ at $V_{GS}=0.5V$ and $V_{DS}=0.5V$) are plotted in Fig. 6 for the different nanowire cases at different channel lengths. It can be clearly seen that the optimum device performance in the ballistic limit happens at a relatively heavier CB minima mass. This emanates from the trade-off condition of utilizing a heavy transport mass. A light transport mass ($m_{t}$) will lead to a high injection velocity or mobility ($\propto 1/\sqrt m_{t}$) . However, at the same time due to the degraded SS the threshold voltage increases, reducing the gate-overdrive impacting the final ON state current. Along the same arguments, a heavy transport mass can lead to gains in term of SS, but the gains do not translate into an improved ON state performance due to a lower injection velocity. It is also expected for a heavy transport mass to lend some advantage in improving the inversion layer charge because of higher density of states or improved $C_{q}$ \cite{Fischetti2007}. A proper definiton of the inversion charge, however, becomes a debatable topic at these ultra-scaled channel lengths hence this is not discussed in detail.

  It is observed that as the channel length ($L_{g}$) is scaled from 7 nm to 3 nm (Fig 6 (a-c)) the CB minima mass at which the peak performance is obtained also increases. This happens because of degraded SS due to reducing gate control with channel length scaling. This means that an even heavier CB minima mass is required to offset the loss of gate control and to improve device performance. The advantage of a heavy effective mass is highlighted for the shortest channel length of $L_{g}=3$ nm . The ON state current continues to improve monotonically with increasing CB minima mass, owing to improving SS (Fig. 6(d)). Engineering a channel material with a transport mass heavier than $\sim$0.95 $m_{0}$ could possiby further enhance the performance at $L_{g}=3$ nm.

After computing the $I_{DS}-V_{GS}$ in the ballistic , the ON state current was recalculated in the presence of electron-phonon scattering \cite{OMENphonon}. The ON state current limited by electron-phonon scattering was computed for  Si$<$100$>$ under the condition of no-stress and a compressive uniaxial stress of -2GPa. It should be noted Si$<$100$>$ are also the best performing nanowire cases in the ballistic limit at the different channel lengths. The phonon limited ON current for InAs is assumed to be $96\%$ of the ballistic ON current  \cite{iedm2010}. Phonon scattering lowers the $I_{ON}$ with Si$<$100$>$ operating at $63\%$ and compressively stresses Si$<$100$>$ operating at $50\%$ of the ballistic limit. It is expected for the compressively stressed Si$<$100$>$ to have a lower ballisticity ratio because of increased density of states for scattering owing to its heavier CB minima mass. However, with reducing channel length the ballistic ratio (B.R.) improve as the distance over which a carrier can back-scatter reduces \cite{CJeong2009}. At $L_{g}=3$ nm,  $Si<100>$ begins to operate at $81\%$ while compressively stresses Si$<$100$>$ operates at nearly $69\%$ of the ballistic limit (Fig 6(c)).

\begin{figure*}[t]
\centering
\includegraphics[scale=0.9]{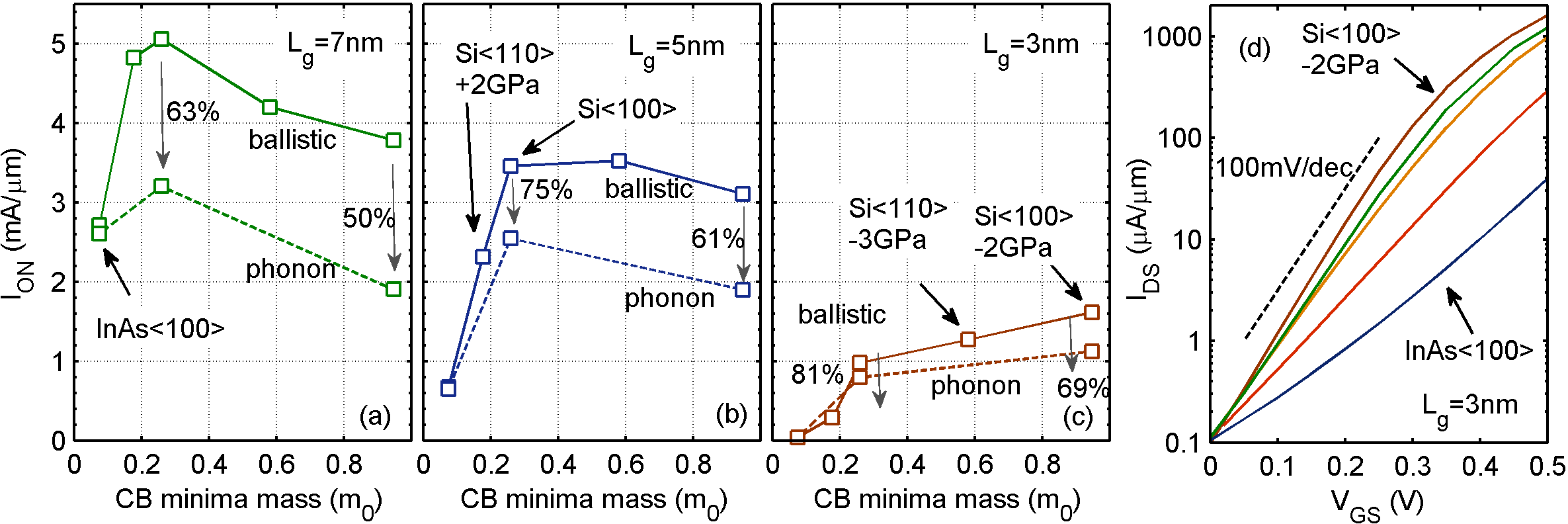}
\caption{ ON state currents in ballistic and phonon scattering limited regime for (a) $L_{g}$=7 nm (b) $L_{g}$=5 nm and (c)  $L_{g}$=3 nm channel lengths. Also the ballisticity ratio is computed for phonon scattering limited ON state current values. (d) $I_{DS}$ vs. $V_{GS}$ plot for the different nanowire cases at $L_{g}$=3 nm in the ballistic limit.} 
\end{figure*}

Interestingly, InAs $<$100$>$, which is currently an actively researched material for future MOSFET scaling, and the tensiled strained Si $<$110$>$, which is the current industry standard, both lag behind in performance at $L_{g}$$<$8 nm. This is a critical understanding that as the channel lengths scale below 8 nm we enter a new regime of device operation where the device performance is severely limited by S-D tunneling. However, at the same time engineering a relatively heavier transport mass can still lead to performance improvements. It is still encouraging to see that the compressively stressed Si$<$100$>$ nanowire with the channel length scaled to 3 nm, even in presence of phonon scattering, is still able to deliver a respectable $I_{ON}/I_{OFF} >10^{4}$ at supply voltage of $V_{DD}=0.5V$ (Fig. 7). 

\begin{figure}[t]
\centering
\includegraphics[scale=0.9]{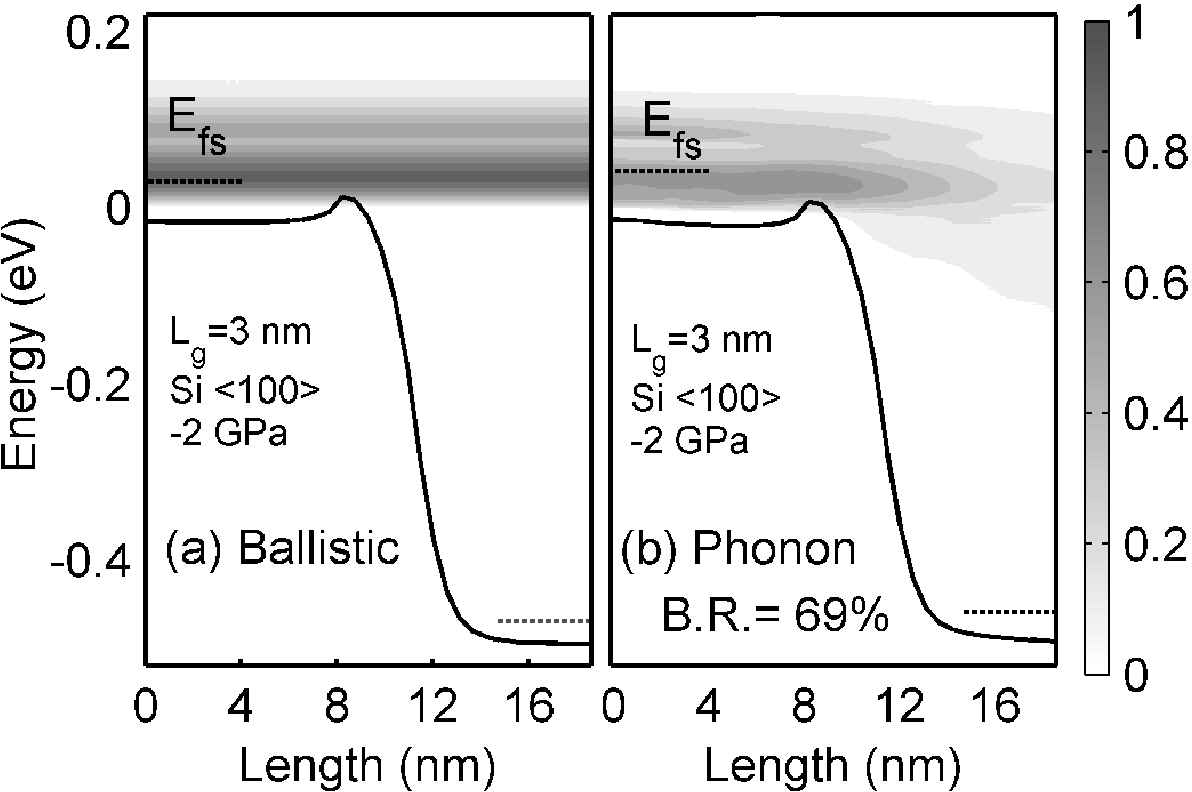}
\caption{ Normalized energy current spectrum for Si$<$100$>$ -2GPa case at $L_{g}=3$ nm in (a) ballistic and (b) phonon scattering limited regime at the ON state.} 
\end{figure}

 Although, the ON state currents are shown to reduce with channel lengths, the design solution to scaling MOSFETs at $L_{g}$$<$8 nm could lie in vertically stacked multiple nanowire MOSFETs that increases the final $I_{ON}$ without compromising on subthreshold characteristics \cite{Bernard2009}.

\section{Conclusion}
In this work the deleterious effects of S-D tunneling in nanowire n-MOSFETs at $L_{g}$$<$8 nm are highlighted in realistically gated and extended devices. Since S-D tunneling depends on the transport mass, as a first step the different CB minima masses that can be engineered in Si and InAs are shown. Next, using full-band quantum simulations based on $sp^{3}d^{5}s^{*}$ tight-binding model the subthreshold and ON state performances are analyzed. The monotonic improvement in subthreshold slope with  heavier CB minima mass is clearly observed as a heavy mass limits S-D tunneling. At the same time the ON state performance does not improve monotonically but shows a peak-like nature. This is due to the tradeoff condition that a heavy (or light) mass offers in terms of OFF-state and ON-state properties. The present results show that the optimal device performance at $L_{g}$$<$8 nm lie in heavier ($>0.25m_{0}$) transport mass channel designs. A realistic scenario of scaling MOSFETs well below  $L_{g}$$<$8 nm could lie in channel designs with a heavy transport mass to improve sub-threshold characteristics and multiple vertically stacked nanowires to boost the ON state current.

\section*{Acknowledgment}
The authors would like to thank Materials, Structures and
Devices Focus Center, which is one of the six research centers
funded under the Focus Center Research Program (a Semiconductor
Research Corporation entity); nanoHUB for the computational
resources; and Rosen Center for Advanced Computing,
National Institute for Computational Sciences, National Center
for Computational Sciences and Texas Advanced Computing
Center for the supercomputer resources.

%The authors acknowledge the support of the MSD Focus Center, one of six research centers funded under the Focus Center Research Program (FCRP), a Semiconductor Research Corporation entity. The use of nanoHUB.org computational resources operated by the Network for Computational Nanotechnology funded by the US National Science Foundation under grant EEC-0228390 is gratefully acknowledged.

\ifCLASSOPTIONcaptionsoff
  \newpage
\fi

\end{document}